\documentclass[aps,prl,epsfigg,twocolumn,showpacs]{revtex4}
\usepackage{ulem}
\usepackage{dcolumn}
\usepackage{bm}
\usepackage{graphicx}
\usepackage{amsmath}
\usepackage{latexsym}
\usepackage{amsfonts}
\usepackage{amssymb}
\usepackage{array}
\begin{document}

\title{Entanglement reciprocation between qubits and continuous variables }

\author{Jinhyoung Lee$^1$, M. Paternostro$^2$, M. S. Kim$^2$, and S. Bose$^3$}

\affiliation{$^1$Department of Physics, Hanyang University,
Sungdong-Gu, 133-791, Seoul and Quantum Photonic Science Research Center, Hanyang University, Seoul 133-791, Korea\\
$^2$School of Mathematics and Physics, Queen's
University, Belfast BT7 1NN, United Kingdom\\
$^3$School of Physics and Astronomy, Gower Street, London
WC1E 6BT, United Kingdom}

\date{\today}

\begin{abstract}
  We investigate how entanglement can be transferred between qubits and
  continuous variable (CV) systems. We find that one ebit borne in 
  maximally entangled qubits can be fully transferred to two CV systems
  which are initially prepared in pure separable Gaussian field with
  high excitation. We show that it is possible, though not straightforward, to retrieve the
  entanglement back to qubits from the entangled CV systems. The
  possibility of deposition of multiple ebits from qubits to the initially unentangled CV systems is
  also pointed out.
\end{abstract}

\maketitle

Quantum information processing (QIP) has been extensively studied
for a qubit system which is a quantum extension of a bit, spanning
two-dimensional Hilbert space.  A qubit is realized by a spin, a
two-level atom, the polarization of a photon and a superconductor
among others.  A two-dimensional system is mathematically handy and
logically easy to treat.  On the other hand, many
continuous-variable (CV) physical systems such as a harmonic
oscillator and a light field, which are defined in
infinite-dimensional Hilbert space, have also attracted considerable
attention for other practical reasons. While qubit and CV systems
are nearly always treated separately, there is a good reason to
believe that a study of their interface may result in synergy for
the implementation of the QIP. There have been some pilot works on
how to entangle two separate qubits by an entangled Gaussian
field~\cite{reznik,kim,cirac}. In this paper, we ask the interesting
questions of how easy it is to: a) deposit the entanglement of two
qubits to a pool of coherent states and b) retrieve quantum
entanglement back to qubits from the pool.

When two maximally entangled two-level atoms are sent to two
respective cavities, initially prepared in vacuum, after the Rabi
time the maximal entanglement is fully transferred to the cavity
fields~\cite{kim01,vanenk}. Here the interaction is assumed resonant
and the cavities are lossless. Essentially, in the above transfer
the cavity does not behave as a true CV system, as only the
$|0\rangle$ and $|1\rangle$ states of cavity take part. The cavity
initially in the vacuum is like a pool without a drop of water so
that dropping a tiny bit of water will be noticeable. However, if
the pools are full of water, an additional drop of water will not
make a difference. This means that there is a chance that when the
cavities are prepared with coherent fields of large amplitudes,
atoms' depositing extra excitation will probably not make a big
difference and the cavity fields will not be entangled much (or even
at all). How about a possibility to retrieve the quantum
entanglement by the second set of atoms which interact with the
cavity fields left by the first set of atoms? Will the atoms be able
to recover the entanglement deposited by the first set of atoms? In
this paper, we find an answer to these questions.

{\it Model}.- Let us consider two atoms in the triplet state
\begin{equation}
|\psi(0)\rangle_{a}={1\over\sqrt{2}}\left(|e\rangle_1|g\rangle_2+|g\rangle_1|e\rangle_2\right),
\label{atom-0}
\end{equation}
where $|e\rangle$ and $|g\rangle$ stand for the excited and
ground states of the atom. This state is maximally entangled and is said to carry one {\it ebit} of entanglement. The two atoms enter their respective
cavities which are initially prepared with coherent states.  For
convenience, we assume that the amplitudes of the coherent states
are same to $\alpha$. The initial state for the atoms and fields
is $|\Psi(0)
\rangle_{af}=|\psi(0)\rangle_a|\alpha\rangle_1|\alpha\rangle_2$.

We consider how much the atomic entanglement, in units of ebit, is transferred to the
infinite-dimensional cavity fields by their resonant interaction.  Under the rotating wave approximation, the interaction Hamiltonian $\hat{H}=\hbar\lambda(\hat{a}^\dag|g\rangle\langle e|+\hat{a}|e\rangle\langle g|$).  The bosonic creation and annihilation operators are denoted by $\hat{a}^\dag$
and $\hat{a}$, respectively, the coupling between the field and
the atom by $\lambda$ and $t$ is the duration of interaction.  In
this case, the evolution of the atom and field state is determined by
the following propagation operator: $\hat{U}=\hat{U}_1\otimes
\hat{U}_2$ where, in atomic atomic bases $\langle e|=(1,0)$ and
$\langle g|=(0,1)$,
\begin{equation}
\hat{U}_i=\left(\begin{array}{cc}
\hat{U}_{11}^{(i)} & \hat{U}_{12}^{(i)} \cr
\hat{U}_{21}^{(i)} & \hat{U}_{22}^{(i)} \cr
\end{array}\right)
\label{evolution}
\end{equation}
with the operators~\cite{phoenix}
\begin{equation}
\label{evol-op}
\begin{aligned}
\hat{U}_{11}^{(i)}&=\cos \lambda t\sqrt{\hat{a}_i\hat{a}_i^\dag},\hskip0.6cm
\hat{U}_{12}^{(i)}=-i\hat{a}_i\frac{\sin\lambda t\sqrt{\hat{a}_i^\dag
    \hat{a}_i}}{\sqrt{\hat{a}_i^\dag\hat{a}_i}},\\
\hat{U}_{21}^{(i)}&=-i\hat{a}_i^\dag\frac{\sin\lambda t\sqrt{\hat{a}_i
    \hat{a}_i^\dag}}{\sqrt{\hat{a}_i\hat{a}_i^\dag}},\hskip0.6cm
\hat{U}_{22}^{(i)}=\cos \lambda t\sqrt{\hat{a}_i^\dag\hat{a}_i}
\end{aligned}
\end{equation}
where the subscript of $\hat{a}$ and $\hat{a}^{\dag}$ denotes the mode of the field.

After the interaction, the atom-field state evolves to $\hat{U}|\Psi(0)\rangle_{a-f}$.
Here, we postselect the cavity field conditioned on two atoms
leaving the cavities in their ground states.  The main reason of
the postselection is to bring the cavity field to a pure state,
whose measure of entanglement is the von Neumann entropy of the
reduced density operator.   While in this paper we are interested
in a possibility for qubits to deposit one complete ebit to a
large CV system, there is no measure or criterion of entanglement
for a general CV state.  The field state after postselection is
\begin{eqnarray}
|\psi(1)\rangle_f =\frac{\cal N}{\sqrt{2}}\left(
\hat{U}_{21}^{(1)}\hat{U}_{22}^{(2)}+\hat{U}_{22}^{(1)}\hat{U}_{21}^{(2)}
\right)|\alpha\rangle_1|\alpha\rangle_2
\label{cavity-field}
\end{eqnarray}
The normalization constant is denoted by ${\cal N}$ and the
coherent state is expanded \cite{barnett} such as
$|\alpha\rangle=\sum_mC_m|m\rangle$ where $C_m
=\alpha^m\mbox{e}^{-{1\over2}|\alpha|^2}/ \sqrt{m!}$ gives
a Poissonian weight with the average photon number $\bar{n}=|\alpha|^2$.
Substituting these into Eq.~(\ref{cavity-field}), we find
\begin{equation}
|\psi(1)\rangle_f =\sum_{n, m=0}^{\infty} {\cal C}_{n,m}|n\rangle_1|m\rangle_2
\label{cavity-1}
\end{equation}
with ${\cal C}_{n,m}=\frac{-i{\cal N}{\mbox e}^{-|\alpha|^2}}{\alpha\sqrt{2}}
 [\frac{\alpha^{n+m}\sin(\lambda
 t\sqrt{n})\cos(\lambda t\sqrt{m})}{\sqrt{m!(n-1)!}}+n\!\leftrightarrow\!{m}]$.

{\it Entanglement transfer from qubits to CV}.- The atoms initially have one ebit as they are maximally
entangled.  We like to know how much ebit is
transferred to the cavity fields by the resonant interaction.  As
the cavity fields are in a pure state $|\psi(1)\rangle_f$, the
amount of ebit ${\cal E}$ is calculated by ${\cal
E}=-\mbox{Tr}\hat{\rho}_{f1}\log_2\hat{\rho}_{f1}$ where the
reduced density operator for the cavity field 1 is
\begin{equation}
\hat{\rho}_{f1}=\mbox{Tr}_{f2}\hat{\rho}_f
=\sum_{m,n,n'}{\cal C}_{n,m}{\cal C}^*_{n',m}|n\rangle\langle n'|.
\label{reduce}
\end{equation}
In Fig.1 {\bf (a)} we plot ${\cal E}$ against $\alpha$ and the
interaction time $\lambda t$ (unit of $\pi$). When $\alpha=0$, we
know ${\cal E}=1$ for sure. Fig.1 {\bf (b)} shows that the probability
of the atoms leaving the cavities in the ground states. When
$\alpha<1$, an oscillating behavior is observed in the degree of
entanglement as well as in the atomic population. On the other
hand, it is interesting to note that when $\alpha$ is large the
cavities are with complete ebit whenever the atoms leave the
cavities in their ground states except the first moments of
oscillations. We can analyze this by showing that
$|\phi_1\rangle\equiv\hat{U}_{21}|\alpha\rangle$ is orthogonal to
$|\phi_2\rangle\equiv\hat{U}_{22}|\alpha\rangle$ in
Eq.~(\ref{cavity-field}). In other words
\begin{equation}
v_0\equiv|\langle\phi_1|\phi_2\rangle| =
\frac{\mbox{e}^{-|\alpha|^2}}{2}\sum_{n}\frac{\sqrt{n}}{\alpha}
\frac{\alpha^{2n}}{n!}\sin 2\lambda t\sqrt{n} \label{ortho1}
\end{equation}
has to be zero. If so, state (\ref{cavity-field}) becomes a maximally entangled qubit state as it will be an equally weighted
superposition of two orthogonal composite states. For simplicity,
let us take $\alpha$ real. In the limit of $\alpha^2\gg 1$, the Poissonian distribution is replaced by a Gaussian distribution over the variable $n$ with mean value and variance equal to $\alpha^2$~\cite{barnett} so that
\begin{eqnarray}
  \label{eq:pdtogd}
  C_n^2 \equiv \mbox{e}^{-\alpha^2} \frac{\alpha^{2n}}{n!} \approx \frac{1}{\sqrt{2\pi
  \alpha^2}} \mbox{e}^{- \frac{(n-\alpha^2)^2}{2\alpha^2}}.
\end{eqnarray}
Taking into account the largely contributing terms of
$\sqrt{n}$, {\it i.e.} those of $n$ near the peak $\alpha^2$, we have
$\sqrt{n} = \sqrt{\alpha^2 + (n-\alpha^2)} \approx \alpha \left(1 +
  \frac{n-\alpha^2}{2\alpha^2}\right)$.
Finally, the summation over $n$ is replaced by an integration in terms of
$x=(n-\alpha^2)/\alpha$ and the integration region is extended
to $(-\infty, \infty)$.  We
then immediately recognize $v_0$ as a Fourier transformation of a
Gaussian function:
\begin{equation}
v_0\propto (\sin2\alpha\lambda t+{\lambda t\over
2\alpha}\cos2\lambda\alpha t)\mbox{e}^{-\frac{\lambda^2t^2}{2}},
\label{approx}
\end{equation}
which decreases exponentially to zero and the two states become
orthogonal to each other exponentially
with regard to the interaction time.  This shows the transfer of a complete ebit from two qubits to a
CV system of a large amplitude.  It is straightforward to show that the probability of the postselection is 1/4 for the limit considered here.

Using the same analogy to prove their orthogonality, we can show that
$\langle\phi_0|\phi_0\rangle=\langle\phi_1|\phi_1\rangle$ for $\alpha\gg1$.
Suppose the initial atomic state was prepared not in the perfect
triplet state (\ref{atom-0})  but in a partially entangled mixed or pure
state.  If we again assume the case of postselecting atoms in their  ground
states, from the earlier analysis we know that the atom initially in
$|e\rangle$ will take the initial coherent field to $|\phi_1\rangle$
and $|g\rangle$ to $|\phi_2\rangle$.  As the two field state bases are
orthogonal with the same weight, it is straightforward to show that the
field state collapses to the state which bears the same amount of
entanglement as in the initial atomic qubits.  This  shows the perfect
transfer of initial entanglement to a CV system.

\begin{figure}[b]
{\bf (a)}\hskip4.0cm{\bf (b)}
\centerline{\includegraphics[width=0.23\textwidth]{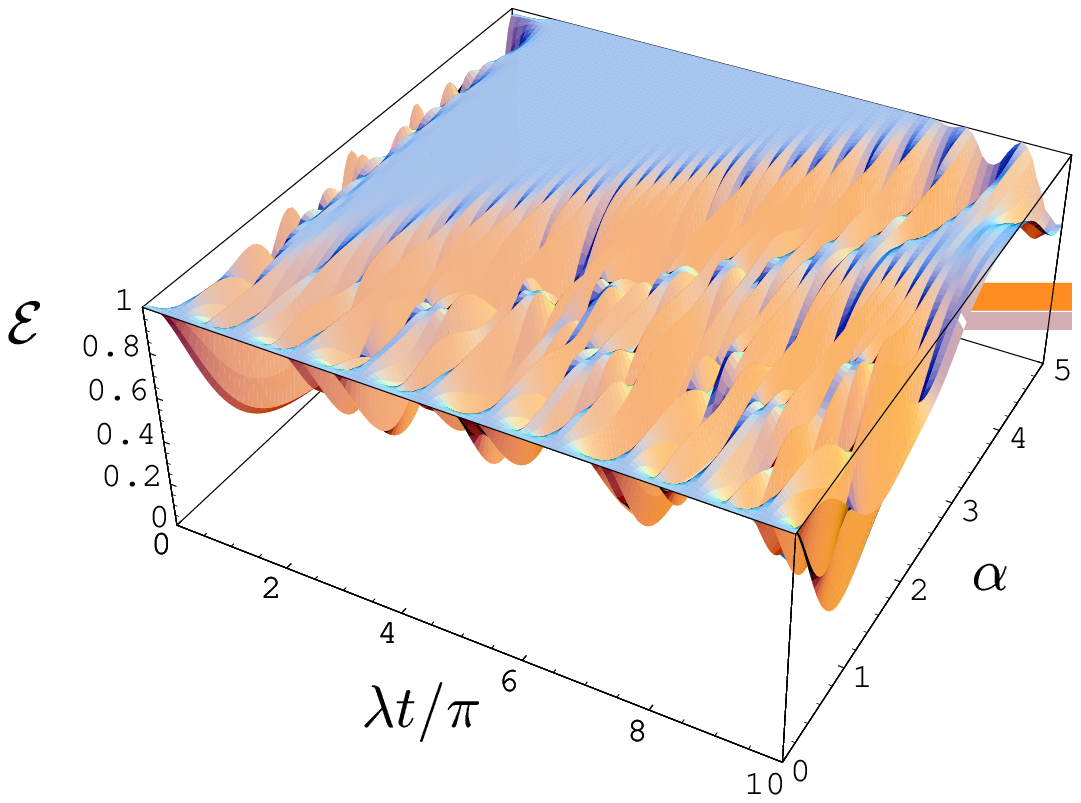}
\hspace*{0.2cm}\includegraphics[width=0.23\textwidth]{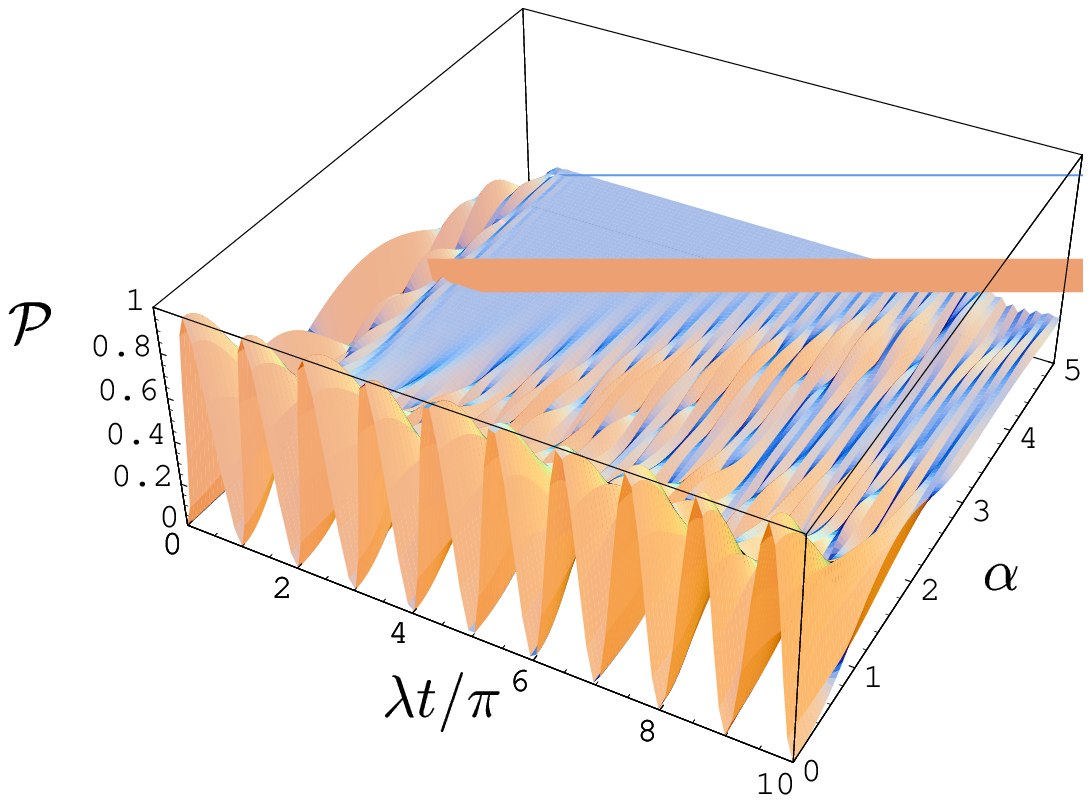}}
\caption{{\bf(a)}: Degree of entanglement for the cavity field
depending on the interaction time $\lambda t$ (in unit of $\pi$) and the amplitude
$\alpha$ of the initial coherent state; {\bf (b)}: Probability of
finding the atoms leaving the cavities in their ground states.
}
\label{fig:fig1}
\end{figure}

In order to see the transfer of the ebit, we took a limit to
ignore the discrete nature of photons. However, it is interesting
to note that we need to recover the discrete nature to explain
the revival of the oscillatory behavior in entanglement as shown
in Fig.~1(a).  The revival occurs when the sinusoidal functions
in Eq.~(\ref{ortho1}) are in phase.  The significant contributions
of the sinusoidal functions come from around the peak of the
Poissonian distribution.  At the peak of the revival time $t_r$:
$2\lambda t_r\sqrt{\alpha^2}-2\lambda
t_r\sqrt{\alpha^2-1}=2\pi$. Taking only the first two terms of
the binomial expansion of the square root, we find the revival
time $t_r=2\alpha\pi/\lambda$.  In fact, the dynamics of
entanglement follows the well-known argument for the collapse and
revivals of Jaynes-Cummings (JC) model~\cite{knight}.

\begin{figure}[htbp]
\centerline{\includegraphics[width=0.4\textwidth]{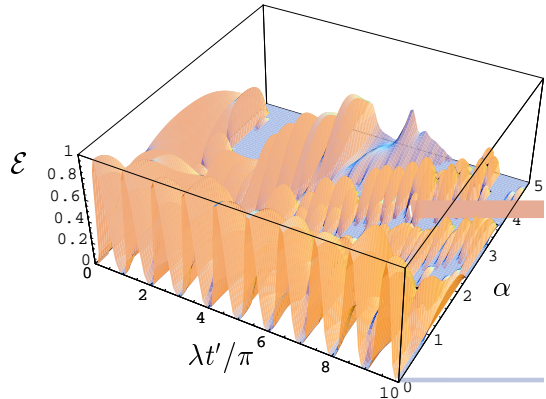}}
\caption{Degree of entanglement for the second pair of atoms depending on the
  interaction time $\lambda t'=\lambda t$ and $\alpha$. 
  }
\label{fig:fig2}
\end{figure}

{\it Entanglement retrievals}.- We have seen that the
qubits can transfer a complete ebit to a CV system conditionally.  The
next question is 'Will it be possible for the qubits to retrieve the
ebit from the CV system?'  In order to solve this problem, we take the
second set of atoms initially prepared in their ground states to send
through the respective cavities which are in $|\psi(1)\rangle_f$.
According to the earlier discussions on the propagation of the
atom-field state, after the interaction time of $t'$, the atom-field
state becomes
\begin{equation}
|\Psi(2)\rangle_{a-f}=\left(\begin{array}{c}
\hat{U}_{12}^{(1)}\hat{U}_{12}^{(2)}\cr
\hat{U}_{12}^{(1)}\hat{U}_{22}^{(2)}\cr
\hat{U}_{22}^{(1)}\hat{U}_{12}^{(2)}\cr
\hat{U}_{22}^{(1)}\hat{U}_{22}^{(2)}\cr
\end{array}\right) |\psi(1)\rangle_f =\sum_{n,m=0}^{\infty}{\mathbf V}_a |n\rangle_1|m\rangle_2
\label{cavity-2-0}
\end{equation}
with the matrix
\begin{equation}
{\mathbf V}_a=\left(\begin{array}{c}
-\sin(\lambda t'\sqrt{n+1})\sin(\lambda t'\sqrt{m+1}){\cal C}_{n+1, m+1} \cr
-i\sin(\lambda t'\sqrt{n+1})\cos(\lambda t'\sqrt{m}){\cal C}_{n+1, m} \cr
-i\cos(\lambda t'\sqrt{n})\sin(\lambda t'\sqrt{m+1}){\cal C}_{n, m+1} \cr
\cos(\lambda t'\sqrt{n})\cos(\lambda t'\sqrt{m}){\cal C}_{n, m} \cr
\end{array}\right).
\label{atomic-matrix}
\end{equation}
In order to investigate how much of the field entanglement deposited by
the first set of atoms, would be transferred to the second set, we trace
$|\Psi(2)\rangle_{a-f}$ over the field variables and find the state of
the atoms:
$\hat{\rho}_a=\sum_{n,m}{\mathbf V}_a{\mathbf V}_a^\dag$.
The degree of entanglement for the two atoms is found using the log
negativity\cite{leekimpark}  of the partial transposition of the density operator $\hat{\rho}_a$
and plotted in Fig.~2 as a function of
the interaction times $\lambda t'$. 
It is seen
that, for non-vanishing $\alpha$, the CV fields are not able to transfer the complete ebit to the
atoms.  However we cannot simply say that it is possible to transfer an
ebit from a qubit system to a CV system while the converse is not true.
The reason is that the qubit$\rightarrow$CV transfer of an ebit was
conditioned on the qubits having lost their entanglement completely.  It
is not straightforward to find such the condition on the CV state for
the CV$\rightarrow$qubit transfer.

\begin{figure}[htbp]
\centerline{\includegraphics[width=0.4\textwidth]{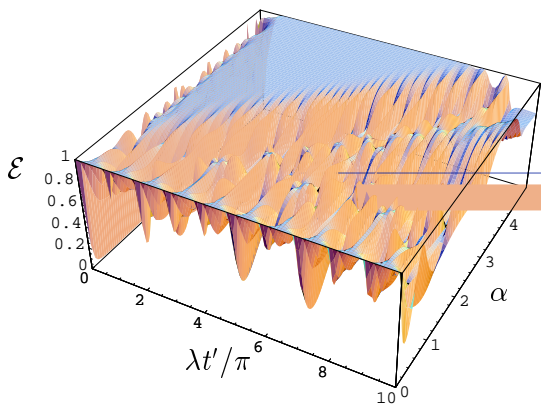}}
\caption{Degree of entanglement for the second pair of atoms at
  the postselected event of $(\alpha, \alpha)$, depending on the
  interaction time $\lambda t'=\lambda t$ and $\alpha$. Note that the
  shape resembles very much Fig.1(a), the degree of entanglement for the
  cavity field.}
\label{fig:fig3}
\end{figure}

In order to improve the degree of entanglement transferred to the atoms,
we consider an orthogonal
  measurement of $\{\hat{P}_\alpha^{(i)}, \hat{Q}_\alpha^{(i)} = \openone -
\hat{P}_\alpha^{(i)}\}$, where $\hat{P}_\alpha^{(i)}$ is the projection onto the
coherent state of its amplitude $\alpha$.  In Fig.~3, the degree of
entanglement for the atoms is plotted, conditioned on the fields in 
$\hat{P}_\alpha^{(1)}\hat{P}_\alpha^{(2)}$ for interaction times $t^\prime=t$.  
We can see the complete entanglement 
transfer for a CV system to a qubit system.   This is analyzed as follows.  By postselecting the event of
$(\alpha, \alpha)$ after the interaction time $t^\prime=t$, the atomic
state becomes
\begin{eqnarray}
  \label{eq:psas}
  |\psi(2)\rangle_a = \frac{{\cal N}^\prime}{\sqrt{2}}
  \begin{pmatrix}
    2 v_1 v_2 \\
    v_1 v_4 + v_2 v_3 \\
    v_3 v_2 + v_4 v_1 \\
    2 v_3 v_4
  \end{pmatrix},
\end{eqnarray}
where $v_1 = \langle \alpha | \hat{U}_{12} \hat{U}_{21} | \alpha
\rangle$, $v_2 = \langle \alpha | \hat{U}_{12} \hat{U}_{22} | \alpha
\rangle$, $v_3 = \langle \alpha | \hat{U}_{22} \hat{U}_{21} | \alpha
\rangle$, $v_4 = \langle \alpha | \hat{U}_{22} \hat{U}_{22} | \alpha
\rangle$. ${\cal N}^\prime$ is the new normalization factor.  Using the same
approximation leading to Eq.~(\ref{approx}), we find that
$v_{1,4}\approx \frac{1}{2} [\pm 1+\cos\left(2 \alpha \lambda t\right)
e^{-\frac{(\lambda t)^2}{2}}]$, where $-$ sign is for $v_1$ and $+$ for
$v_4$, and $v_2= v_3\approx {\cal O}(t)e^{-\frac{(\lambda t)^2}{2}}$
where ${\cal O}(t)$ is a linear sum of sinusoidal functions.  In the long time
limit $\lambda t\gg 1$, $v_2=v_3\rightarrow 0$ while $v_1=-v_4=-{1\over
  2}$.  State (\ref{eq:psas}) is now a maximally entangled triplet
state (\ref{atom-0}), which has been perfectly retrieved after the
interactions with the cavity fields. This can be inferred from the analysis of the entanglement between the second pair of atoms shown in Fig.~(\ref{fig:fig3}). It is surprising to note that the
probability of getting the coherent state is as high as
  50\%, in this limit.

It is worth stressing that the results presented in this work are a feature of the coherent state used as a memory for entanglement. Indeed, we have checked that, by considering initial thermal states for the fields and applying the protocol in the previous paragraphs, the initial ebit in the qubit-state can never be completely deposited in the CV state and, consequently, retrieved from it. Obviously, in order to investigate the entanglement deposit, one faces the hard problem of quantifying the entanglement in a two-mode non-Gaussian state. This difficulty has been bypassed adopting the technique described in ref.~\cite{sougato} based on the projection onto a subspace spanned by the bidimensional bases $\{|{n}\rangle, |{n+1}\rangle\}_{j}$ with $j=1,2$. The entanglement within the resulting projected state is then averaged over thermal weighting functions characterized by their mean photon number $\bar{n}$. This provides us with a lower bound to the entanglement in the two-mode non-Gaussian state. It is thus straightforward to see that the perfect deposit-retrieval process is possible just for the trivial case of $\bar{n}=0$ for the initial fields.

An interesting question to ask now is what happens when a series of
atom pairs, each in the state $\psi(0)\rangle_a$, are allowed to
interact with the cavity fields (in the usual setting of one atom with
each cavity field). We found that, for example, for   $\alpha= 4.5$ and
if the first pair of atoms had interacted for a time $t_1 =
6.47/\lambda$  (which deposits an ebit of entanglement), then a second
pair of atoms interacting for a time $t_2 = 11.04/\lambda$ deposits
another ebit, and a third pair interacting for a time $t_3 =
3.24/\lambda$ deposits yet another ebit. Each of these depositions have
a success probability of   $\sim 0.25$ and are robust to small
variations in $t_i$, as before. This contrasts the case of the cavities
starting in vacuum states where incommensurate Rabi frequencies prevent
the deposition of more than one ebit through the resonant interaction.
The cavities in our case can thus serve as "stationary" reservoirs for
multiple ebits supplied by atom pairs in the form of "flying" qubits,
which may be difficult to hold in other situations. In addition, these
multiple ebit entangled cavity states may be directly used for
teleportation of higher dimensional states. Application of
$\hat{P}_\alpha$, $\hat{Q}_\alpha$ also allows the retrieval of 1.82
ebits and 1.91 ebits at optimized times from the 2- and 3-ebit entangled
cavity states respectively through a pair of atoms.


   As a final remark, we would like to shortly point out that our
approach is quite setup-independent. Obviously, an implementation
based on quantum electrodynamics in cavity would be the most natural
choice~\cite{harochekimble}. However, the interaction model we have
assumed, the resonant JC one, turns out to be naturally valid in
many physical situations in which coherent exchange of excitations
between spin-like particles and bosons are
involved~\cite{wilsonrae}. The very recent progress in micro and
nano-fabrication of integrated cavity-qubit systems in the
semiconductor and superconducting domain~\cite{schoelkopf} and the
readily available sources of coherent states in many ranges of
frequency makes our proposal adaptable to different physical
situations. The language we have adopted in this paper, thus, has to
be seen as a pure matter of convenience.



{\it Remarks}.- In this paper, we have considered interface between two
hetero-dimensional systems.  An ebit can be transferred to a CV system
from a qubit system and back to the qubit system conditionally. One
extremely nice thing is that the transfer happens in the quasi-steady
state, which means that one does not have to be careful in picking the time for entanglement transfer.  We also found an interesting analogy between the entanglement
reciprocation and the collapse and revival of Rabi oscillations in the
JC model considered for the proof of discreteness in the
photon number.  The perfect reciprocation of entanglement is a
particular feature of a coherent state.  Postselecting the fields in $(\alpha,\alpha)$ is not trivial.  Even though a heterodyne detection or a beam splitter detection may approximate it, this deserves a further investigation.

\acknowledgments



\begin{thebibliography}{99}
\bibitem{reznik}B. Reznik, quant-ph/0008006 (2000).
\bibitem{kim} W. Son {\em et al},
\jmo {\bf 49},
1739 (2002); M. Paternostro {\em et al},
\prl {\bf 92}
197901 (2004); M. Paternostro {\it et al.}, \pra {\bf 70} 022320 (2004).
\bibitem{cirac} B. Kraus and J. I. Cirac, \prl {\bf 91}, 013602 (2004); A. Retzker, J. I. Cirac and B. Reznik, \prl {\bf 94}, 050504
(2005).
\bibitem{kim01} M. S. Kim, presented at the ``100 Years Werner Heisenberg – Works and Impact''
(26-30 September 2002).
\bibitem{vanenk} S. J. van Enk, quant-ph/0507189 (2005).
\bibitem{phoenix} S. J. D. Phoenix and P. L. Knight, Ann. Phys. {\bf
    186}, 381 (1998).
\bibitem{barnett} S. M. Barnett and P. M. Radmore, Methods in
    Theoretical Quantum Optics (Oxford, 1997).
\bibitem{leekimpark} J. Lee, M. S. Kim, Y. J. Park and S. Lee,
\jmo {\bf 47}, 2151 (2000).
\bibitem{knight} Jaynes and Cummings, Proc. IEEE {\bf 51}, 89 (1963); B. W. Shore and P. L. Knight, \jmo {\bf 40}, 1195 (1993).
\bibitem{sougato} S. Bose, I. Fuentes-Guridi, P. L. Knight, and V. Vedral, Phys. Rev. Lett. {\bf 87}, 050401 (2001).
\bibitem{harochekimble} J. M. Raimond, M. Brune, and S. Haroche,
Rev. Mod. Phys. {\bf 73}, 565 (2001).
\bibitem{wilsonrae} M. Paternostro, G. Falci, M. S. Kim, and G.M. Palma, \prb {\bf 69}, 214502 (2004); I. Wilson-Rae and A. Imamoglu, {\prb} {\bf 65}, 235311 (2002);  F. Marquardt and C. Bruder, {\prb} {\bf 63}, 054514 (2001).
\bibitem{schoelkopf} G. S. Solomon, M. Pelton, and Y. Yamamoto,
Phys. Rev. Lett. {\bf 86}, 3903 (2001); F. Jelezko {\it et al.}, Phys. Rev. Lett. {\bf 93}, 130501 (2004); A.Wallraff {\it et al.}, Nature (London) {\bf 431}, 162 (2004).


\end{thebibliography}
\end{document}